\documentclass{elsart}
\usepackage{graphicx}
\usepackage{graphics}
\usepackage{amssymb}
\usepackage{amsmath}
\usepackage{bm}

\newcommand{\be}{\begin{equation}}
\newcommand{\ee}{\end{equation}}
\newcommand{\bea}{\begin{eqnarray}}
\newcommand{\eea}{\end{eqnarray}}
\newcommand{\beal}{\begin{align}}
\newcommand{\eal}{\end{align}}
\newcommand{\bespl}{\begin{split}}
\newcommand{\espl}{\end{split}}

\newcommand{\nslash}{\kern 0.2 em n\kern -0.50em /}
\newcommand{\kslash}{\kern 0.2 em k\kern -0.45em /}
\newcommand{\pslash}{\kern 0.2 em p\kern -0.50em /}
\newcommand{\Sslash}{\kern 0.2 em S\kern -0.50em /}
\newcommand{\Pslash}{\kern 0.2 em P\kern -0.50em /}
\newcommand{\Rslash}{\kern 0.2 em R\kern -0.50em /}

\begin{document}

\begin{frontmatter}

\title{A nuclear-style mean field 
model for the Sivers effect in nucleon-hadron single spin 
asymmetries and Drell-Yan. 
} 

\author{A.~Bianconi}
\address{Dipartimento di Chimica e Fisica per l'Ingegneria e per i 
Materiali, Universit\`a di Brescia, I-25123 Brescia, Italy, and\\
Istituto Nazionale di Fisica Nucleare, Sezione di Pavia, I-27100 Pavia, 
Italy}
\ead{andrea.bianconi@bs.infn.it}

\begin{abstract}
I study the effect of scalar and spin-orbit absorption 
potentials, in the 
production of a nonzero Sivers-like asymmetry 
in hadron-hadron high energy collisions 
(Drell-Yan and single spin asymmetries). 
A basic 
model is built for the intrinsic state of a 
quark in the projectile hadron. S-wave and P-wave 
2-component states 
are considered. 
Before the hard event, this 
quark 
is subject to absorbing mean fields 
simulating interactions with a composite target. 
The relevant 
interaction terms are found to be the imaginary diagonal spin-orbit 
ones. 
Spin rotating terms, and scalar absorption, seem not to be 
decisive. 
For $x$ $=$ 0 the found Sivers 
asymmetry vanishes, while at larger $x$ its qualitative 
dependence on $x$, $K_T$ 
follows the usual trends met in available models and parameterizations. 
Given the present-day knowledge of the considered phenomenological 
interactions, it is not possible to establish whether the related 
Sivers-like asymmetry is a leading twist-one.  
\end{abstract}

\begin{keyword}
High energy hadron-hadron scattering, Sivers asymmetry, spin-orbit 
forces. 
\PACS 13.85.Qk,13.88.+e,13.90.+i 
\end{keyword}

\end{frontmatter}

\maketitle

\section{Introduction}

\subsection{General background}

The problem of the study and measurement of T-odd distributions 
in hadron-hadron 
scattering has recently acquired 
a certain relevance and quite a few related experiments 
have been thought or scheduled for the next ten 
years\cite{panda,assia,pax,rhic2,compassDY}. 

In particular several studies and models have been proposed 
for the Sivers distribution function\cite{Sivers}. Its possible 
existence as a leading-twist distribution was 
demonstrated\cite{BrodskyHwangSchmidt02,Collins02,
JiYuan02,BelitskyJiYuan03} recently, and 
related\cite{JQVY06} to previously 
studied T-odd mechanisms\cite{EfremovTeryaev82,QiuSterman91}. Some 
phenomenological forms for its dependence on $x$ and $K_T$ have been 
extracted\cite{Torino05,VogelsangYuan05,CollinsGoeke05,BR06a} 
from available data\cite{Star,Hermes,Compass,Phenix}. 

While studies of general 
properties\cite{Pobylitsa03,DalesioMurgia04,Burkardt04,
Drago05,GoekeMeissnerMetzSchlegel06,BoerVogelsang06,Entropy3} 
of T-odd functions 
relate these functions with a wide spectrum of phenomena,  
quantitative models mostly follow the general scheme suggested 
in \cite{BrodskyHwangSchmidt02}. A known quark-diquark 
spectator model\cite{JakobMuldersRodriguez97} is extended by 
including single particle (meson or gluon) 
exchange\cite{BoerBrodskyHwang03,GambergGoldsteinOganessyan03,
BacchettaSchaeferYang04,LuMa04}. In the case of \cite{Yuan03} the 
unperturbed starting model was a Bag model. 

\subsection{This work}

The class of processes I want to consider here is the one of 
single spin asymmetries in collisions between an unpolarized 
hadron and a transversely polarized proton. In particular, 
azimuthal asymmetries in 
Drell-Yan dilepton production and hadron semi-inclusive 
production, where in both cases one of the colliding hadrons is 
normal-polarized. I consider phenomena that may be 
present in the beam energy range 10-300 GeV (so, not necessarily 
leading twist effects). 

The present work uses 
phenomenological schemes that are not typical of 
perturbative QCD, but rather of high-energy nuclear 
physics. It 
is inspired by previous works on T-odd 
structure functions in high-energy nuclear physics\cite{BB95,BR97}, 
by a previous work on nuclear-target induced polarization in 
Drell-Yan\cite{BR_JPG}, and by the 
results from the theory and phenomenology 
of spin-orbit interactions in high-energy hadron-hadron 
exclusive processes 
(see e.g. \cite{Akchurin89,BKLST99,Crabb90}
and references therein). 

The goal is not to reproduce precisely 
some phenomenology. 
Rather, it is to establish whether scalar and spin-orbit interactions 
dominated by their absorption part are 
able to build a nonzero Sivers-like asymmetry with a reasonable shape, 
possibly of higher-twist nature. 

For this reason, the model for 
both the initial ``intrinsic'' state of a quark in the proton, 
and for initial state interactions 
is built in such a way to be 
as simple as possible. All the necessary functions have been 
chosen in Gaussian form and the parameter number has been reduced 
to the minimum necessary to explore the interesting 
independent physical situations. 

\subsection{The general scheme}

Contrary to the ordinary treatment of the problem, where one 
works on a two-point correlation operator deriving from 
a set of squared one-point amplitudes,  
I develop most of the work at 
the level of one-point amplitudes, 
square them and then sum over 
the relevant states. I imagine, for a hadron with a given 
spin projection $S_y$ $=$ $+1/2$, a two-component quark 
spinor $(f_+,f_-)$, 
with $f_\pm$ associated with the transverse quark spin.  
In this scheme, the $\gamma_+-$trace normally calculated on 
the correlation operator simply corresponds to the sum 
$|f_+|^2+|f_-|^2$. This quantity is the final goal of the 
calculation, and the 
distribution functions associated to an unpolarized 
quark, including Sivers' one, are extracted from it. 

Concerning the initial, ``intrinsic'' state of a quark 
in the hadron,  
I assume that the relevant quantity defining this state 
is the quark total angular momentum $\vec J$ in the hadron rest frame. 
In this state a nonzero correlation 
$<\vec S_{hadron} \cdot \vec J_{quark}>$ $\sim$ $+1/4$ is present. 
In other 
words, the quark $\vec J$ coincides with the 
parent hadron spin. This may be realized both in S-wave and in 
P-wave, with spin-spin correlation 
$<\vec S_{hadron} \cdot \vec S_{quark}>$ $\sim$ $\pm1/4$.
A nonzero correlation between the hadron and the 
quark angular momentum is necessary in any model, since a spin-related 
effect is impossible 
if a quark transports no information on the parent hadron spin. Clearly, 
we have different effects in the S and P-wave cases. 

This correlation alone would not produce 
a single-spin asymmetry of naive Time-odd origin because of global 
invariance rules. 
Initial state interactions between 
the two hadrons, or between the quarks of one hadron and the quarks of 
the other one, must be introduced\cite{BrodskyHwangSchmidt02,Collins02}. 

I reproduce these  interactions in eikonal approximation 
$exp(\int \hat T d\xi)$, where $\hat T$ is a 2x2 space-time 
dependent matrix reproducing a mean field acting on the projectile quark, 
and $\xi$ is a light-cone coordinate. 
Then the full matrix element, 
affected by this operator, is calculated 
in space-time representation. 

Although it may seem more natural to adopt a mean field 
treatment for problems where a projectile is subject to multiple 
soft scattering\cite{Feshbach}, there is also a 
tradition\footnote{the 
most obvious reference is \cite{Glauber59}, but see also 
ref.\cite{Anisovich} 
for an exposition of the application of these techniques 
to quark scattering on composite hadronic structures in light-cone 
formalism.} 
for such recipes in 
hadronic problems dominated by a single or at most 
double hard scattering event, when 
the scatterer belongs to a composite 
structure

The way it is used here, the above eikonal approximation is only 
a short-wave approximation, that alone does not support the 
persistence of initial state interaction effects at very large 
energies. This derives from the fact that we do not know 
the asymptotic properties of the operator $\hat T$ (see below). 
In addition, the above eikonal approximation 
does not support automatically factorization, since it is applied 
at single point amplitude level. 

\subsection{Anti-hermitean initial state interactions}

The added initial state interactions (the $\hat T$ matrix) 
consist of two terms: 
(i) anti-hermitean scalar mean field, (ii) anti-hermitean 
spin-orbit mean field.\footnote{In  
the following the words ``real'' and ``imaginary'' are sometimes 
used instead of ``hermitean'' and ``anti-hermitean'' when speaking of 
operators. 
}

Hermitean terms have been tested and they 
affect the results. Alone, these terms do 
not produce T-odd distributions 
(by definition, since they they are intrinsically 
T-even) and do not change the main qualitative features of 
the presented results.  
Not to overload this work with a many-parameter 
phenomenology, I have limited myself to terms that  
are not intrinsically T-even. 

A remark is important: strong and electromagnetic interactions 
are T-even and hermitean. As well known in nuclear 
physics\cite{Feshbach}, relevant anti-hermitean terms originate 
in the projection of hermitean interactions on a subspace, 
including only a part of all those degrees of freedom that are able 
to exchange energy/momentum within a characteristic interaction 
time relevant for the problem. If one were able to include 
all the relevant degrees of freedom in the formalism, there would 
be no room for anti-hermitean interactions (see 
\cite{Entropy3} for a long discussion about these points). 

\subsection{Spin-orbit terms}

The results of this calculation show that 
also most of the considered anti-hermitean terms  
are not effective, for the purpose of a Sivers asymmetry. 
The key interaction term is one of the 
three components of the spin-orbit scalar product. 
A chain of qualitative arguments presented in 
section III relates imaginary spin-orbit terms to the 
high-energy hadron-nucleon analyzing power and recoil 
polarization. 
These observables are nonzero 
at as large beam energies as 300 GeV
(for the case of 
quasi-forward scattering\cite{Akchurin89} ) 
or 25-30 GeV (for the case of large transferred 
momenta up to 7 GeV/c, see e.g.\cite{Crabb90}). Their behavior 
at larger energies is not known, and there is no 
commonly accepted model that allows for an 
extrapolation\cite{Akchurin89,BKLST99,Crabb90}. 

This has two consequences. On the one side, at energies 
$\sim$ 100 GeV we face the possibility of relevant Sivers-like 
asymmetries with this origin. On the other side, it 
is impossible to decide whether this is a 
leading twist effect. For this reason, as above anticipated, 
it is impossible to decide whether the $\hat T$ operator appearing 
in the rescattering factor $exp(\int \hat T d\xi)$ is 
nonzero at very large energies. Because of this, 
in the 
following the terms ``Sivers asymmetry'' and ``Sivers effect'' 
are preferred to 
``Sivers function''. The latter is appropriate in the case of 
a leading twist contribution. Experimentally, it may 
be impossible to distinguish between the two at the presently 
available energies. 

\section{The general formalism}

Where not differently specified, all variables will refer 
to the center of mass of the colliding hadrons. Let 
$\vec b$ $=$ $(b_x,b_y)$, 
be the quark impact parameter and 
$\vec K_T$ $=$ $(k_x,k_y)$  
the transverse momentum conjugated 
with it. Let $P_+$ be the large light-cone component of the 
hadron momentum, so that $xP_+$ is the quark $(+)$ momentum 
conjugated with $z_-$. 

I substitute $z_-$ with the rescaled coordinate 
\begin{equation}
\xi\ \equiv\ P_+z,\hspace{0.5truecm} 
\rightarrow \hspace{0.5truecm}
P_+ \int dz_- exp(-ixP_+z_-)\ =\ 
\int d\xi exp(-ix\xi) 
\label{eq:fourier0}
\end{equation}
not to work with a singularity of the 
Fourier transform 
in the infinite momentum limit $P_+$ $\rightarrow$ $\infty$. 

Since the inclusive process is described here in terms of squared 
amplitudes, and these amplitudes are calculated before being 
squared, $\xi$ is not bound 
to be positive, as it happens in the ordinary treatment based on a 
two-point correlator with intermediate real states. In that 
case $\xi$ has the meaning of the difference between 
the light-cone positions of two points. 
Here it describes the light-cone position of one of the two only. 

\subsection{Basic structure of the quark unperturbed state and 
insertion of initial state interactions} 


I represent the initial 
``unperturbed'' quark state in the form

\begin{eqnarray} 
\vec \psi(\xi,\vec b)\ 
\equiv\ 
\int dx d^2K_T \vec f(x,\vec K_T) e^{ix\xi} e^{i\vec K_T \cdot \vec b}, 
\hspace{0.5truecm}
\vec f(x, \vec K_T)\ \equiv\ 
\left(
\begin{array}{cc}
f_+(x,\vec K_T) \\
f_-(x,\vec K_T) 
\end{array}
\right)
\label{eq:fourier1} 
\end{eqnarray} 

So our hadron consists in a coherent superposition of plane wave 
states with given $x$, $\vec K_T$ and transverse spin, each with amplitude 
$f_+(x,\vec K_T)$ or $f_-(x,\vec K_T)$. 

I suppose that the parent hadron has 
$y-$polarization $+1/2$, and that 
one initial state only contributes to the final distribution 
function. The expected distribution  
has the form\footnote{This 
definition is the one given by the 
so-called ``Trento convention''\cite{Trento} for polarization 
oriented as written. 
It assumes that 
the second term is scale-independent and in this case $q_S$ is 
the Sivers function. 
Since this work refers to energies $\sim$ 10$\div$300 GeV, 
I will speak of ``Sivers asymmetry'' referring to the full second term 
$q_S K_x/M$. 
}

\begin{equation}
q(x,\vec K_T)\ =\ |f_+(x,\vec K_T)|^2\ +\ |f_-(x,\vec K_T)|^2\ 
\equiv\ q_U(x,K_T)\ +\ {K_x \over M} q_S(x,K_T).
\label{eq:sivers}
\end{equation} 

The Sivers asymmetry can of course be isolated by subtracting 
two terms like the previous one, corresponding to opposite 
hadron polarizations. Here I limit myself to searching for 
$k_x-$asymmetric terms in the above $q(x, \vec K_T)$ 
unpolarized quark distribution 
corresponding 
to one assigned hadron transverse polarization. 

To introduce initial state interactions, I identically write 
$\vec f(x,\vec K_T)$ as a twice iterated Fourier transform, and 
in the intermediate stage I substitute 
each plane wave spinor by a spinor that contains 
the distortion due to the initial state interactions. 
Writing only the $x,\xi-$dependence 
for simplicity, 
it means that in the undistorted plane wave 

\begin{equation}
\vec f_{PW}(x)\ \equiv\ 
\int d\xi e^{-ix\xi} \int dx' e^{ix'\xi} \vec f_{PW}(x')
\ \equiv\ 
\int d\xi e^{-ix\xi} \int dx' \big[ e^{ix'\xi} \hat I \big]\vec f_{PW}(x') 
\end{equation}
(where $\hat I$ is the 
identity matrix and 
in the last passage I have only highlighted the piece to be modified) 
the free field operator $exp(ix'\xi) \hat I$ 
is substituted by the more general matrix operator 
$\hat \Psi(x',\xi)$ reproducing a field subject to the action 
of initial state interactions: 

\begin{equation}
\vec f_{PW}(x)\ 
\rightarrow\ \vec f_{DW}(x)\ \equiv\ 
\int d\xi e^{-ix\xi} \int dx' \hat \Psi(x',\xi) \vec f_{PW}(x')
\end{equation}

More precisely, initial state interactions 
in eikonal approximation affect the quark light-cone path  
starting from $\xi$ $=$ $-\infty$ and reaching the hard interaction 
point $\xi$, 
along fixed impact parameter lines (see the discussion 
in refs.\cite{Collins02} and \cite{Burkardt04}, and compare the 
figures 
describing final state interactions for Deep Inelastic Scattering 
in ref.\cite{BrodskyHwangSchmidt02} with  
those for initial state interactions in Drell-Yan in 
ref.\cite{BoerBrodskyHwang03}). 

These initial state interactions are here averaged by an 
effective mean field 
containing absorbing and spin orbit terms. So, each 
plane wave is substituted by a wave with eikonal phase 
distorted by this local field: 

\begin{eqnarray}
e^{ix\xi} e^{i\vec K_T \cdot \vec b} 
\left(
\begin{array}{cc}
f_+ \\
f_- 
\end{array}
\right)
\ \rightarrow\ 
e^{ix\xi} e^{i\vec K_T \cdot \vec b} 
exp\Bigg(
\int_{-\infty}^\xi \hat T(\xi',\vec b) d\xi'
\Bigg)\cdot 
\left(
\begin{array}{cc}
f_+ \\
f_- 
\end{array}
\right)
\label{eq:DW1}
\end{eqnarray}
where $\hat T$ is a 2x2 matrix operator. 

\subsection{The undistorted quark state}

In this subsection I 
refer the quark spin, orbital and total angular momentum 
to the parent hadron rest frame. 

In absence of initial state interactions, we may assume that we 
are able to calculate the Fourier transform eq.(\ref{eq:fourier1}) 
and write it directly in impact parameter representation 
as ($PW$ means ``plane wave'')

\begin{equation}
(\ref{eq:fourier1})_{PW}\ =\ 
\vec \psi(\xi,\vec b)\ \equiv\ \phi(\xi)\phi'(|b|)
\cdot|J_y\ =\ +1/2> 
\label{eq:groundstate0}
\end{equation}

\noindent 
where my main interest is for the very simple $S-$wave state 

\begin{eqnarray}
|J_y\ =\ +1/2>_S\ \equiv\ 
\left(
\begin{array}{cc}
1 \\
0 
\end{array}
\right)
\label{eq:groundstate1s}
\end{eqnarray}

\noindent 
and as a second choice for the P-wave state 
\begin{eqnarray}
|J_y\ =\ + 1/2>_P\ \equiv\ 
ib_x\ 
\left(
\begin{array}{cc}
0 \\
1 
\end{array}
\right).
\label{eq:groundstate1p}
\end{eqnarray}

Eq.(\ref{eq:groundstate0}) reproduces a space-time fluctuation 
of the hadron ground state into a quark+spectator state. 
Eqs.(\ref{eq:groundstate1s}, \ref{eq:groundstate1p}) 
are the impact parameter space 
projections of the states 
$Y_{00} |1/2>_y$, 
$Y_{11}(\theta_y,\phi_y) |-1/2>_y$. 
Of course, other terms may be included. I limit to these two 
possibilities. 

The above states eqs. (\ref{eq:groundstate1s}) and 
(\ref{eq:groundstate1p}) imply a positive/negative 
transversity (since 
these states transport information on the polarization of the 
parent hadron) 
but do not allow for any 
$\vec K_T-$odd asymmetry, including a Sivers asymmetry,  
in absence of initial state interactions. 

Both the above states imply $J_y$ $=$ $+1/2$ for the quark and 
the parent hadron in the hadron rest frame. This is the 
limit possibility. 
More in general we may imagine that a state where the parent 
hadron is fully polarized with transverse spin $S_y$ $=$ $+1/2$,  
corresponds to a quark mixed configuration of the kind  
\begin{equation}
a\ \Big| |J_y\ =\ 1/2> \Big|^2\ +\  
b\ \Big| |J_y\ =\ -1/2> \Big|^2. 
\end{equation} 

\noindent
For $a$ $\approx$ $b$ evidently the quark transports 
little or no information on 
the parent hadron polarization state, so it is logically impossible to 
get a nonzero hadron-polarization-related function, unless some 
very indirect mechanism is imagined.  
So, in the following I exclude this possibility. 
For nonzero $b$ $\neq$ $a$, 
the results I get must be diluted by the 
factor $|a-b|/(a+b)$. Indeed, substituting the quark 
$|J_y\ =\ 1/2>$ initial state 
with the quark $|J_y\ =\ -1/2>$ initial state, the 
asymmetries that are calculated in the following 
simply reverse their sign. 

For $\phi(\xi)$ and $\phi'(b)$, 
I simply take Gaussian shapes $exp(-y^2/{y_o}^2)$. The same 
is done for the 
relevant functions $\rho(\xi)$ and $\rho'(b)$ later introduced to 
describe initial 
state interactions. 
In practice, the underlying 
hadron-quark-spectator vertex is a space-time version of the 
vertex adopted in \cite{GambergGoldsteinOganessyan03} (a Gaussian 
quark-diquark vertex). 

The parameters for these gaussian functions  
are chosen not to get too unrealistic distributions. Since 
for $x$ $=$ 0 the distribution is large, the above must 
be considered an implementation of a sea+valence state. 
This is not a decisive detail, 
because the used initial state interactions lead to 
a zero Sivers effect at $x$ $=$ 0.  
For $x$ $=$ 1 the parameters are tuned so to have a small 
distribution value, that cannot be zero however. With a 
logarithmic mapping it could be possible to produce a distribution 
that is zero at $x$ $=$ 1. This increase of complication would not 
be worthwhile since this work does not focus on the $x$ $\approx$ 1 
region where a completely different physics should be included. 

\subsection{Initial state interactions} 

I assume that the distorting factor 
$\hat D$ $\equiv$ $exp[\int d\xi \hat T(\xi,\vec b)]$ of eq.(\ref{eq:DW1}) 
does not depend on $x$ or $\vec K_T$. This simplifies much the 
calculations since it allows for transporting $\hat D$ out of  
the Fourier transform eq.(\ref{eq:fourier1}) and applying directly 
it to the function 
$\vec \psi(\xi,\vec b)$ of eq.(\ref{eq:groundstate0}).\footnote{Else, 
one could directly define the action of an $x-$dependent 
initial state interaction starting 
from eq.(\ref{eq:groundstate0}), sacrificing something at interpretation 
level.} 

So, this equation is modified to (DW means ``Distorted Wave''): 

\begin{eqnarray}
(\ref{eq:fourier1})_{DW}\ =\ 
\phi(\xi)\phi'(|b|) \cdot 
exp\Bigg(
\int_{-\infty}^\xi \hat T(\xi',\vec b) d\xi'
\Bigg)\cdot
|J_y\ =\ + 1/2>_{S,P}.  
\label{eq:DW2}
\end{eqnarray} 

In the calculations, the exponential 
operator is approximated by a quasi-continuous product: 

\begin{equation}
\int_{-\infty}^\xi \hat T(\xi',\vec b) d\xi'
\ \approx\ 
\prod (1 + \hat T d\xi)
\end{equation}
where the product starts from a negative and large enough $\xi'$ 
value where interactions may be neglected, and stops at $\xi$. 
The $\hat T$ matrix is 

\begin{eqnarray}
\hat T\ \equiv\ \left(
\begin{array}{cc}
-(\delta + \alpha b_x) &  -i\alpha b_y \\
i\alpha b_y &  -(\delta - \alpha b_x) 
\end{array}
\right)
\rho(\xi)\rho'(b).
\end{eqnarray}

All the coefficients are supposed to introduce reasonably 
small corrections, at least for $K_T$ $\lesssim$ 3 GeV/c where we know 
that any asymmetry due to 
initial state interactions is at most 30 \%. 

The $O(\delta)$ term is a scalar absorption term, associated with 
spreading of the quark momentum and so to damping of the quark initial 
state. It assumes underlying chaotic interactions, that 
because of this lack of coherence deplete any given 
$|x,\vec K_T>$ state without a direct coherent enhancement of 
another one, as it would happen in the case of a hermitean 
interaction. A part of the lost flux is 
recovered because of diffraction, but in the average some flux is lost 
from the elastic channel. 

The $O(\alpha)$ terms 
are spin-orbit terms, since in a basis 
where $S_y$ is diagonal, as the one I are using here, we may write 

\begin{equation}
\hat T d\xi\ =\ \rho\rho' \Big( -\delta \hat I\ -\ \alpha b_x \hat \sigma_y\ 
+\ \alpha b_y \hat \sigma_x \Big) P_+ dz_-  
\label{eq:T1}
\end{equation}

Defining 
\begin{equation}
\delta\ \equiv\ \delta' x/\sqrt{2}
, \hspace{0.5truecm} \alpha\ \equiv\ \alpha' x/\sqrt{2},  
\label{eq:T2}
\end{equation}
and 
remembering that, at large $P_+$, 
in the hadron collision c.m. frame 
\begin{equation}
k_z\ \approx\ xP_+/\sqrt{2},\hspace{0.5truecm}   
L_y\ \approx\ -k_z b_x,\hspace{0.5truecm}   
L_x\ \approx\ k_z b_y,\hspace{0.5truecm}   
L_z\ <<\ L_x,L_y, 
\label{eq:T3}
\end{equation}
(here $\vec L$ is referred to the hadron collision c.m. frame) 
the above may be rewritten as 

\begin{equation}
\hat T d\xi\ =\ 
\rho\rho' \Big(-\delta' k_z\hat I\ -\ \alpha' \vec L \cdot \vec 
\sigma \Big) dz_-  
\label{eq:T4}
\end{equation}
and we see that $\hat T$ contains a scalar absorption term 
plus a 
spin-orbit term. 

Since it appears in a real exponential (without an explicit 
factor ``i'' in the argument),  
for real $\alpha$ the spin-orbit term is a anti-hermitean one. 
For imaginary $\alpha$, it is hermitean. Since by definition the latter 
cannot produce $T-$odd effects, I have focused my attention on 
the case of real $\alpha$. This corresponds to the 
nuclear physics case of 
an imaginary spin-orbit potential. More in general, we will have 
a complex potential, able to introduce $T-$odd effects if its 
imaginary part is nonzero. Aiming at studying the simplest possible case, 
I limit to a pure anti-hermitean term. 

With the parameter values here assumed (see below), the combined 
action of nonzero $\delta$ and $\alpha$ is such as to produce absorption 
through all 
the region affected by serious initial state interactions. This 
absorption is spin-orbit-selective.

The functions $\rho(\xi)$, $\rho'(b)$ have been chosen with gaussian 
form, and their widths satisfy  
the conditions: $\rho(\xi)$ $\approx$ 
$|\phi(\xi)|^2$, $\rho'(b)$ $\approx$ $|\phi'(b)|^2$. This is motivated by 
the following facts: 
(i) initial state interactions cannot take place too 
far from the hard quark-antiquark vertex; (ii) the projectile 
and the target are supposed to have similar shapes; (iii) in terms of the 
longitudinal rescaled quantity $\xi$ $=$ $P_+z_-$, leading twist 
effects (if any) must take place over a finite $\xi$ range 
in the scaling limit $P_+$ $\rightarrow$ $\infty$; if they are 
next-to-leading, at any finite $P_+$ for which they assume a 
non-negligible value they take place over a finite 
(scale-dependent) $\xi$ range; 
(iv) since I assume that initial state 
interactions have incoherent character, the $\phi(..)$ functions 
are wavefunctions, while 
the $\rho(..)$ functions are densities $\sim$ $|\phi(..)|^2$. 

The choice of using all gaussian functions, with correlated 
widths, is aimed to simplicity and to reducing the number of independent 
parameters. 

\section{Some general comments on the use of a non-hermitean spin-orbit 
mean field in hadron-hadron semi-inclusive processes}

Presently, nonzero effects of spin-orbit terms are measured in 
exclusive hadron-nucleon interactions at rather large energies. 
Their origin in terms of fundamental 
interactions is not fully explained. So they must be considered 
phenomenological interactions. 
What is argued in the present work is 
that these interactions are present at quark-quark level, 
where they preserve the same generic structure they have 
at hadron-hadron level. As a consequence, their effect should be visible 
not only in a few exclusive channels, 
but in a wider class of hadron-hadron induced 
processes, including semi-inclusive scattering and 
Drell-Yan. Although
elastic channel measurements are the most precise available, up 
to now spin-orbit effects have been found in any exclusive channel where 
they have been searched for via dedicated 
experiments. 

In the following I give a qualitative account 
(i) of 
the present-day knowledge on spin-orbit interactions in hadron-hadron 
interactions at high energies, 
(ii) of the 
assumptions that are necessary to pass from hadron-hadron spin-orbit 
interactions 
to the quark-mean field interactions introduced in the previous section. 

\subsection{Present status of spin-orbit interactions in exclusive 
hadronic processes} 

The arguments summarized in this subsection may be reconstructed from 
refs.\cite{Akchurin89,BKLST99,Crabb90}, that also contain 
a long list of references  
concerning theoretical models and previous measurements. Several 
papers on this subject, or touching this subject, have been written 
especially in the years before 1985. 

Writing the amplitude for the elastic scattering between a 
normally polarized beam particle 
and an unpolarized target particle as 
$A(\vec K_T)$ $\equiv$ $A_{even}(\vec K_T)$ $+$ 
$A_{odd}(\vec K_T)$,  
the spin-orbit potential is the core part of 
the impact parameter space representation of $A_{odd}(\vec K_T)$. 
In other words, 
a nonzero normal spin analyzing power in hadron elastic scattering 
on hadron targets is equivalent to a 
spin-orbit coupling like the one appearing in eq.(\ref{eq:T1}) 
of the previous section. As a specific example, 
from eqs.(49) and (53) of ref.\cite{BKLST99} one may directly deduce 
the spin-orbit terms 
in eq.(\ref{eq:T1}) of the present work, taking into account that 
in the case of a Gaussian 
density one has $\partial \rho(b^2) /\partial b_x$ 
$\propto$ $-b_x \rho(b^2) $ (in \cite{BKLST99}, the same 
2-dimensional formalism used here is employed). 

Experimentally, 
an unexpectedly large normal analyzing power (or, equivalently,  
a large recoil polarization)  
is found in nucleon-hadron exclusive processes like elastic scattering. 
In the quasi-forward diffraction-dominated region, this is nonzero 
up to beam energies 300 GeV. At large transferred momenta (up to 7 
GeV/c) it is rather large up to beam energy 30 GeV. What happens 
at larger energies is not known, and it has been guessed but not 
demonstrated that such terms can survive 
asymptotically.\footnote{The 
case of the electromagnetic-strong interference at small angle 
must be excluded from this discussion. It may be asymptotically relevant 
without implying conclusions about spin-orbit 
terms of strong origin. 
}
The separation between the two regimes (small and large 
transferred momenta) is not obvious in the data. 
Comparing small and large angle data (that normally refer to 
different energy regimes) 
taking as a starting point  
the peak at very small angles due to electromagnetic-strong interference, 
and then increasing the angle, 
the transferred momentum dependence of the 
analyzing power or recoil polarization shows 
a series of diffraction or interference peaks and 
changes of sign. The peaks do not decrease in magnitude, and 
at the largest transferred momenta the size of the effect seems to 
increase 
in an uncontrollable way, together with error bars. 
I remark that the region that theoretically should 
imply the transition between the two 
regimes (transferred momenta between 0.5 and 3 GeV/c) 
is also the most important for a nonzero and measurable Sivers 
asymmetry. 

At intermediate energies 
the origin of the measured spin-orbit terms may be reasonably interpreted 
for pion-proton diffractive 
elastic scattering. In this case two dominating Regge trajectories 
(pomeron and rho) mix, leading to interference between 
two terms with phase difference 90$^o$ in the helicity-flip and 
helicity-non-flip amplitudes. 
In the case of quasi-forward elastic proton-proton scattering, several 
Regge poles and cuts potentially contribute and the situation is less 
clear. 
Because the presence of the discussed observables requires 
a phase difference 90$^o$ between 
the interfering helicity-flip and helicity-non-flip amplitudes, 
contributions from $two$ poles/cuts are needed to explain a 
nonzero analyzing power or recoil polarization. So, an unexpected 
survival of these observables at large energies would contradict the 
standard idea that only pomeron contributes asymptotically to 
small-angle elastic scattering. 

For the case of large transferred momentum, there are a lot of 
competing models (e.g. references 5-22 in 
ref.\cite{Crabb90}). In all of them some 
non-trivial mechanism is added to a standard PQCD set of 
processes. Indeed, 
PQCD may be properly applied in this region, leading to zero 
normal analyzing power because of helicity conservation. So, PQCD 
cannot be ignored, and at the same time it cannot be applied 
in the most straightforward way. 

\subsection{Assumptions about 
spin-orbit interactions at quark level}

To pass from the measurements of normal spin asymmetries 
in elastic scattering to the spin-orbit potential contained 
in the previous section, some assumptions are needed. 

Assumption (A): At least a part of the 
high-energy spin-orbit interactions between hadrons 
can be reduced to an incoherent sum of 
spin-orbit interactions between quarks. 

Assumption (B): The absorption part of these quark-quark spin-orbit 
interactions is relevant. 

These two assumptions are plausible and questionable at the same time, 
because of our lack of knowledge about the fundamental mechanisms 
determining the spin-orbit coupling at hadron level. 

In the case of large angle scattering, assumption (A) could be 
naively  
justified because of the short wavelength perturbative regime. 
On the other side, in this regime PQCD may be applied, and it 
says that spin-orbit interactions should 
not exist at all (both at quark and at hadron level), while they are 
there. 

In the case of quasi-forward scattering, we may imagine that 
the spin-orbit term in hadron-hadron scattering is associated with 
an 
interference term between two t-channel Regge pole exchanges. Although it 
is possible to imagine the same process with the same exchanged 
poles taking place between two 
individual quarks $q_1$ and $q_2$, small transverse momenta imply 
a large degree of coherence between two processes like $q_1-q_2$ 
and $q_1-q_3$ scattering. 
So the possibility of extracting the incoherent part of these 
interactions and estimate its relevance would require at least 
to know what exactly is exchanged, and this is not completely clear 
presently. 

Concerning assumption (B), 
it is easy to imagine a large absorption part in quark-hadron 
scattering, for the same reasons 
why hadron-hadron scattering is 
absorption-dominated. 
However, this argument cannot be transferred directly 
from the hadron level to the quark level. 
Indeed, the absorption part of the hadron-hadron 
scattering potential derives from 
the inelasticity of the process. In other words, from the direct 
loss of flux of the initial 
channel.\footnote{the 
word ``direct'' reminds that a part of this lost flux does anyway 
contribute to the elastic channel because of diffraction. 
} 
But what is a hard inelastic process at hadron level may be a 
perfectly elastic process at quark level, in the normally 
employed approximation where quarks are free particles. 

For the specific case of the production of T-odd effects, 
I have long discussed this point 
in my previous work ref.\cite{Entropy3}. Summarizing very briefly, 
since a quark is never 
asymptotically free, inelasticity is not determined by a 
modification of the quark 
internal state or by extra particle production, 
but also by the behavior of the surrounding 
environment. 
This has little relevance in the analysis of T-even effects 
in hard processes, 
because of the separation between hard and soft scales. In T-odd 
effects however, soft scales acquire indirect relevance. In particular, 
elastic quark-quark in-medium scattering leads to a finite imaginary 
part in the quark propagator, i.e. the quark behaves like a free but 
unstable 
particle.\footnote{Of 
course the quark itself is stable from the point of view of its 
internal structure, what is unstable 
is the quark ``free'' state we are 
probing.} 

As a consequence of this inelasticity, if spin-orbit interactions 
may be transferred from hadron to quark level 
an imaginary part is likely to be present, giving an argument 
for the plausibility of 
assumption (B).\footnote{I 
cannot argue that this imaginary part is also dominant, 
as it happens e.g. for the unpolarized part of the forward hadron-hadron 
scattering 
process.
}

\subsection{Mean field potential, exponentiation and asymptotic 
behavior} 

At finite beam energies $\sim$ 10-100 GeV, 
from assumption (A) and (B) the possibility follows of 
writing the cumulative effect of spin-orbit 
quark-quark interactions in the 
exponentiated form eq.(\ref{eq:DW2}) where a quark interacts 
with a mean-field potential along a straight path. 
Exploiting the fact that 
for finite $P_+$, $\xi$ may be one-to-one related in a 
non-singular way to the longitudinal space variable $z$ in the rest 
frame of the unpolarized particle, we may just treat the interaction 
in Glauber-like style\cite{Glauber59} and proceed as in high-energy 
hadron-nucleus collisions. With things done this way, 
exponentiation is just a useful short-wavelength approximation, 
but does not prove that one is considering leading twist 
terms. In other words, I cannot presently support the following 
assumption: 

Assumption (C): The operator $\hat T$ of eq.(\ref{eq:DW1}) is 
finite in the limit 
$P_+$ $\rightarrow$ $\infty$. 

Formal considerations apart, it is obvious that 
such an assumption could make sense if we knew 
the $P_+$ $\rightarrow$ $\infty$ behavior 
of spin-orbit effects in nucleon-hadron elastic 
scattering.  


\section{Results} 

All calculations refer to $k_y$ $=$ 0, so in the following only 
the $k_x-$dependence appears explicitly. 

Figures 1 to 4 have been calculated with the S-wave state, 
figures 5 and 6 with the P-wave one. 

The free parameters have the same values in all figures 1 to 6, 
and their list follows: 

For the gaussian function/density widths: 

$\phi(\xi)$: $\Delta \xi$ $=$ 3.5.  $\rho(\xi)$: $\Delta \xi$ $=$ 2. 
$\phi'(b)$: $\Delta b$ $=$ 0.9. $\rho'(b)$: $\Delta b$ $=$ 0.6. 

I have used 
$\rho(\xi)$ $\approx$ $|\phi(\xi)|^2$ and $\rho'(b)$ $\approx$ $|\phi'(b)|^2$, 
so to have two independent parameters only.  

For the interaction matrix: $\delta$ $=$ 0.2, $\alpha$ $=$ 0.1.  

These mean an overall 20 \% reduction of the quark distribution 
for $K_T$ $=$ 0, entirely due to the parameter $\delta$. At large 
$K_T$ on the contrary the distribution is enhanced. 
The spin-orbit parameter $\alpha$ produces 
local flux modifications with zero $\vec b-$average. The Fourier 
transform for $k_x$ $=$ 0 is not sensible to 
these. For nonzero $k_x$, $\alpha$ produces asymmetry. 


\begin{figure}[ht]
\centering
\includegraphics[width=9cm]{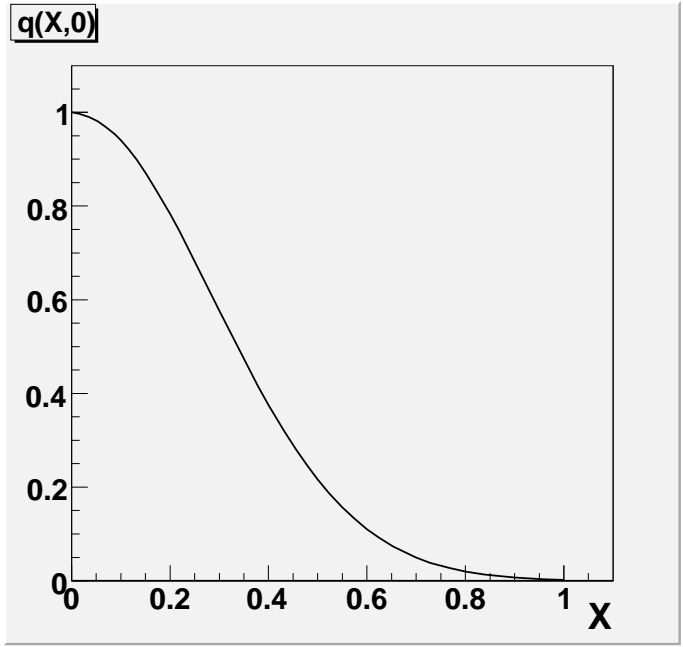}
\caption{S-wave: 
The $q(x,k_x)$ distribution function for $k_x$ $=$ 0, as a function 
of $x$. $k_x$ is the component of $\vec K_T$ orthogonal to the 
initial hadron polarization $\vec S$ $\propto$ $\hat y$. 
For the parameter values, see the beginning of the ``Results'' 
section in the text. 
\label{c0}}
\end{figure} 

\begin{figure}[ht]
\centering
\includegraphics[width=9cm]{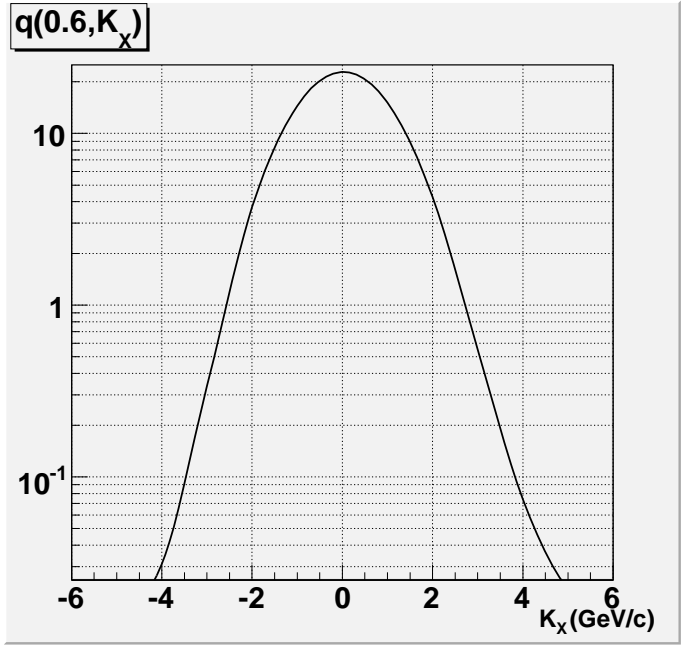}
\caption{S-wave: 
$k_x-$dependence of the quark distribution $q(x,k_x)$ for 
$x$ $=$ 0.6. The asymmetry is made more evident by the logarithmic plot.  
\label{c4}}
\end{figure}

\begin{figure}[ht]
\centering
\includegraphics[width=9cm]{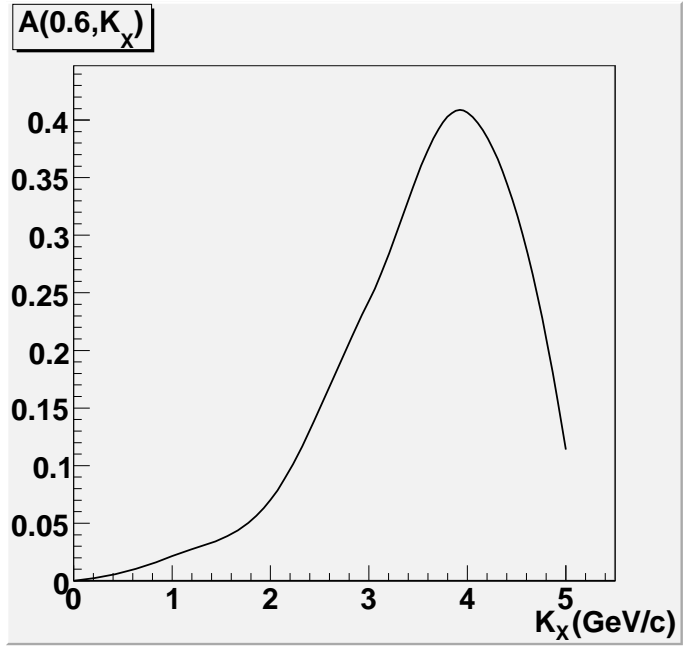}
\caption{S-wave: 
$k_x-$distribution of the asymmetry $(q_+-q_-)/(q_++q_-)$ for 
$x$ $=$ 0.6. It is the left-right asymmetry of the distribution 
reported in the previous figure. 
\label{c2}}
\end{figure}

\begin{figure}[ht]
\centering
\includegraphics[width=9cm]{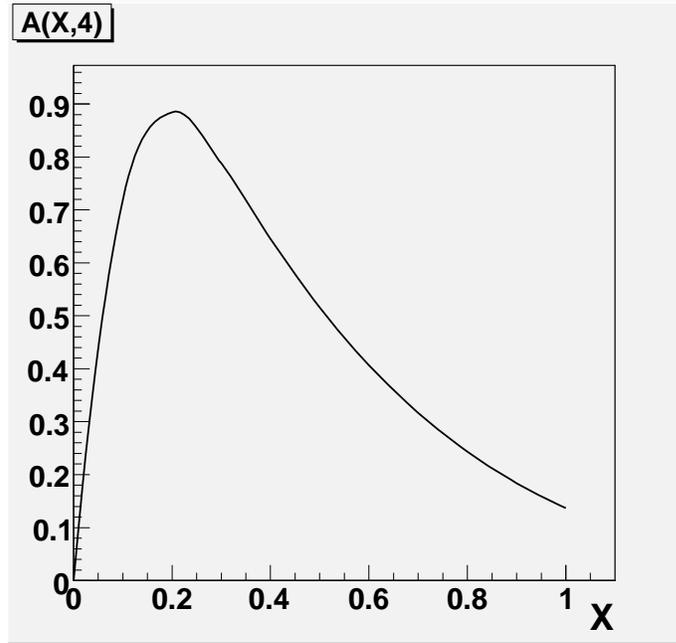}
\caption{S-wave: Asymmetry $(q_+-q_-)/(q_++q_-)$ as a function of $x$ 
for $k_x$ $=$ 4. At this $k_x$ we have 
the peak value of the asymmetry, for each $x$. 
\label{c1}}
\end{figure}

\begin{figure}[ht]
\centering
\includegraphics[width=9cm]{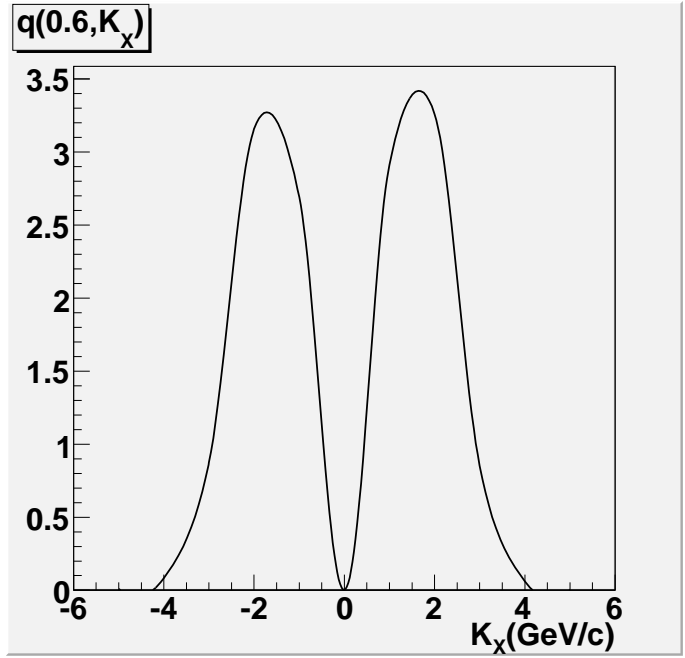}
\caption{
P-wave: 
$k_x-$dependence of the quark distribution $q(x,k_x)$ for 
$x$ $=$ 0.6. 
\label{c4d}}
\end{figure}

\begin{figure}[ht]
\centering
\includegraphics[width=9cm]{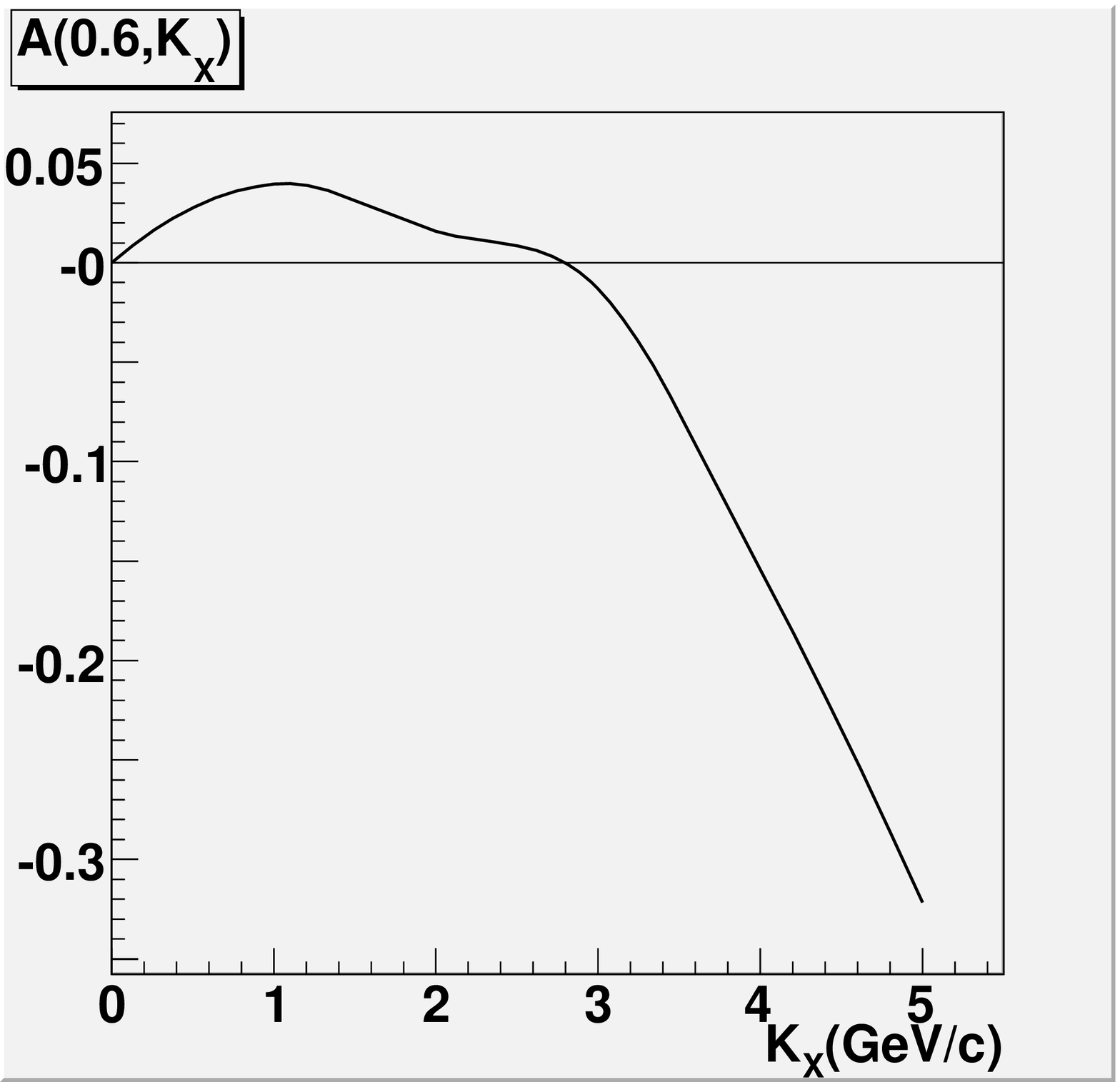}
\caption{P-wave: The asymmetry $(q_+-q_-)/(q_++q_-)$ 
of the quark distribution reported in the 
previous figure. 
\label{c3d}}
\end{figure}

In fig.1 I show the distribution function 
$q(x,k_x)$ for $k_x$ $=$ 0 
in the $x-$range (0,1). 

In fig.2 I show the corresponding 
$k_x-$distribution $q(x,k_x)$ for $x$ $=$ 0.6. A logarithmic plot 
has been chosen to give more evidence to the large$-k_x$ asymmetry, caused 
by a nonzero $\alpha$. 

In fig.3 I show the 
asymmetry $A$ $\equiv$ $(q_+-q_-)/(q_++q_-)$, for $x$ $=$ 0.6,  
as a function of $k_x$ in the $k_x$ range (0,5) GeV/c. With 
$q_\pm$ I mean $q(x,\pm |k_x|)$. 

In fig.4 I show the same asymmetry for fixed $k_x$ $=$ 4 GeV/c 
(i.e. at its peak value) as a function of $x$. 

Many more (not shown) distributions and asymmetries have been produced 
changing the values of all the above parameters. The result is that 
figs.1 to 4 are general enough and contain all the relevant 
qualitative features of the found asymmetries. By tuning parameter 
values,  
the asymmetry may be made larger/smaller, and its peak may be shifted 
towards larger/smaller $x$ or $k_x$.\footnote{I refer to reasonable 
parameter values only. E.g., there is no hint in the literature 
of unpolarized $k_x-$integrated distribution functions being 
strongly affected by 
initial state interactions, so this is considered a boundary 
condition to be respected.}

Summarizing the general properties deduced by systematic parameter 
tuning activity I find: 

1) For small values of the parameters $\delta$, $\alpha$ 
(both $\leq$ 0.2) one already obtains 
peak values of the asymmetry $|A(x,K_{max})|$ $\sim$ 1 
(see fig.4).  This 
is however reached at $k_x$ $\gtrsim$ 4 GeV/c, where experimental data 
would easily present large error bars making it difficult to distinguish 
between e.g. 60 \% and 30 \%  
asymmetries.\footnote{For 
asymmetries over 30 \% the error is given by the error on the 
less populated of the two subsets in comparison. E.g. with total 
population 300 event, 33 \% asymmetry means 100 vs 200 events. 
A fluctuation like 50 vs 250, leading the asymmetry to 
66 \%, is quite easy. In a restricted $x-$ range like $(0.2-0.3)$ in the 
valence region, to 
collect much more than 300 events with $P_T$ $>$ 4 GeV/c is not trivial.
} 
For $k_x$ up to 3 GeV/c the 
asymmetry is smaller than 30 \%, i.e. much smaller than the peak value.  
A consequence of the use of gaussian shapes 
is the presence of this rather pronounced peak at large $k_x$, that 
however 
would have scarce influence on an event-weighted asymmetry (dominated 
by $k_x$ $\approx$ 1$\div$2 GeV/c). 

2) Changing the gaussian distribution parameters it is possible 
to change the shape of the asymmetries, so to have the peak asymmetry 
e.g. at 3 GeV/c. This would be far from anything observed up to 
now (compare with the fits by 
refs.\cite{Torino05,VogelsangYuan05,CollinsGoeke05,BR06a}).  

3) For $k_x$ $>$ 5 GeV/c, we find oscillations near zero, that could be of 
numeric origin. The fast decrease immediately following the peak 
at $k_x$ $=$ 4 GeV/c 
is however stable with respect to changes of the numerical parameters 
(number of integration 
points, integration range cutoffs). 

4) Asymmetries obtained via anti-hermitean spin-orbit terms are zero at 
$x$ $=$ 0. $|A(x,K_{max})|$ increases with $x$ up to a maximum and 
then decreases at larger $x$, seemingly to reach zero at the unphysical 
value $x$ $=$ $\infty$.  
The fact that the asymmetry is zero for $x$ $=$ 0 is a consequence 
of the symmetries of the model, 
and of the fact that the $x$ $=$ 0 component of the Fourier transform 
is just a plain $\xi-$average. 
In a more realistic model this property should be anyway 
present for another reason. Here, we have assumed $x-$independent 
interactions, but accordingly to 
eqs.(\ref{eq:T1}-\ref{eq:T4}), the spin-orbit 
potential is $O(x)$ at small $x$ 
(in a semi-classical approximation, assuming continuous $\vec L$). 
At a semi-classical level, it is reasonable to 
imagine that a wee parton has comparatively small 
$L_x$ and $L_y$ in the hadron collision c.m. frame, and so 
negligible spin-orbit interactions. 

5) Alone, a nonzero 
$\delta$ is not able to produce a Sivers-like  
asymmetry. Nonzero $\alpha$ 
is required. On the other side, $\alpha$ could produce asymmetry 
for $\delta$ $=$ 0 too, 
but this combination ($\delta$ $=$ 0, $\alpha$ $>$ 0) 
would imply an unphysical local increase of particle 
flux. 
With the chosen values of the coefficients we have spin-selective 
absorption, but anyway absorption, with the exception of large$-b$ 
regions where initial interactions are anyway suppressed by $\rho(b)$. 

6) In the P-wave case the $k_x-$event 
distribution assumes the obvious shape shown in fig.5.  
At the 
distribution peaks of fig.5 the asymmetry is $\sim$ 5 \% and has the same 
sign of the S-wave asymmetry of 
figs 1-4. In the distribution tails $k_x$ $>$ 3 GeV/c 
the asymmetry reverses its sign. To analyze in detail situations 
where both waves (S and P) are relevant 
one should however introduce a more realistic different radius for 
the $S-$wave and the $P-$wave distributions. 

7) The non-diagonal terms of the interaction 
matrix eq.(\ref{eq:T1}) 
have practically no role. 
I.e., removing the diagonal $\pm \alpha b_x$ terms from the 
interaction matrix 
produces immediately zero asymmetry, while removing the 
$\pm i \alpha b_y$ terms, or changing their sign, 
only produces negligible changes in the 
asymmetry shape of 
figs. 3 and 4. 

The role of the $\alpha b_x$ terms may be understood 
if one approximates the action of the damping potential in 
a homogeneous form. Then a nonzero $\alpha$ (together with 
a nonzero $\delta$) means that the two components $a_+$ and $a_-$ 
of a spinor $(a_+,a_-)$ are substituted by 
damped exponentials: 

\begin{eqnarray}
\left(
\begin{array}{cc}
a_+  \\
a_- 
\end{array}
\right)
\ \rightarrow\ 
\left(
\begin{array}{cc}
a_+ exp(-\delta+\alpha b_x)\xi \\
a_- exp(-\delta-\alpha b_x)\xi
\end{array}
\right)
\label{eq:damped1}
\end{eqnarray} 

Evidently at large negative values 
of the product $\xi b_x$ we have dominance of the spin state 
$(1,0)$, and the opposite at large positive $\xi b_x$. This asymmetry 
is detected by the $exp(-ik_xb_x)$ Fourier transform. 

Introducing further approximations (substituting 
the gaussian functions with simple cutoffs of the integration ranges, 
and taking $\delta$ $=$ 0) 
shows that the so simplified problem enjoys a kind of $b\cdot \xi$ 
invariance. 
As a consequence, any 
time a parameter is changed so to decrease the $\Delta x$ range 
of the asymmetry, it also increases the corresponding $\Delta k_x$ 
range and viceversa. 
A systematic parameter tuning work confirms that this property 
is approximately present in the full model.  

Some final observations: 

The author of this work has remarked in \cite{Entropy3} 
that damping terms in initial state interactions 
contribute producing T-odd distributions, 
so he cannot claim surprise for the 
results of the present calculation. What is however surprising is 
the non-effectiveness of the 
non-diagonal terms in the $\hat T$ matrix. 
Since one cannot 
exclude that this is due to the exaggerated level of 
symmetry contained in the here used 
unperturbed state, a deeper study needs to be 
devoted to this point. 

Also, one cannot exclude that a more complicated 
structure of initial states may lead to $k_x-$asymmetries in presence 
of $spin-independent$ initial state interactions. This is what 
happened e.g. in refs.\cite{BB95,BR97}, where a T-odd structure 
was produced in presence of a spin-independent (final state) 
interaction.\footnote{The rescattering terms used in those two works 
had a rather different structure and physical meaning each with respect 
to the other. However both are based on non-hermitean scalar interaction 
terms. 
}  
In that case however, the $effect$ of final state interactions was not 
spin independent, despite their hamiltonian was. This was due to 
the elaborate shell-model structure of the initial states. E.g., 
in \cite{BB95} it was possible to get an asymmetry from 
$^{12}$C, but not from $^{16}$O (fully filled 
$P_{3/2}$ and $P_{1/2}$ shells in the latter 
case, only $P_{3/2}$ in the former). 

\section{Conclusions} 

Summarizing, starting from the assumption that the quark 
total angular momentum is 
dominantly oriented as the parent hadron spin, it is possible to 
build a Sivers-like asymmetry via 
mean field initial state interactions of imaginary spin-orbit kind. 
With these interactions, a nonzero Sivers function is present 
also starting from a simple $S_y=1/2$ S-wave ground state 
for the quark. 

Phenomenological interactions of this kind are known in high-energy 
hadron and nuclear physics.
For values of the parameters such as to guarantee a small overall 
effect of initial state interactions ($\approx$ 20 \% distribution 
damping at small $K_T$) the 
qualitative $x,k_x-$distribution of the calculated 
asymmetries follows the typical pattern proposed by widespread 
parameterizations. 

The employed quark ground state has the features of a joint 
valence$+$sea state. Despite this, the predicted asymmetries 
are zero at $x$ $=$ 0. 

In the chosen interaction matrix, the only effective terms are 
the (spin-orbit selective) diagonal absorption ones. In other words, 
spin rotating interactions are not decisive. Also, spin-independent  
absorption alone is not sufficient. 
It is not possible however to exclude that the no-effectiveness 
of these terms is related to the excess of simplicity of the 
initial configuration. 

With the present day knowledge of spin-orbit hadronic interactions, 
it is not possible to establish whether their effects persist at 
very large energies. Consequently it is not possible to establish 
whether a Sivers-like asymmetry generated by them is a leading 
twist one or just an intermediate energy effect. 



\end{document}